# Interference-engineered shortcut to perfect state transfer

Yichuan Zhang[1†], Xuanyu Liu[2†], Zemeng Lin[1], Wange Song[1,2,3*], and Shuang Zhang[1,3,4,5,6*]

*[1]New Cornerstone Science Laboratory, Department of Physics, The University of Hong Kong, Hong Kong, China.*

*[2]National Laboratory of Solid State Microstructures, Key Laboratory of Intelligent Optical Sensing and Manipulations, Jiangsu Key Laboratory of Artificial Functional Materials, College of Engineering and Applied Sciences, Nanjing University, Nanjing, 210093, China.*

*[3]State Key Laboratory of Optical Quantum Materials, The University of Hong Kong, Hong Kong, China.*

*[4]Department of Electronic and Electrical Engineering, The University of Hong Kong, Hong Kong, China.*

*[5]Materials Innovation Institute for Life Sciences and Energy (MILES), HKU-SIRI, Shenzhen, China.*

*[6]Quantum Science Center of Guangdong-Hong Kong-Macao Great Bay Area, 3 Binlang Road, Shenzhen, China.*

*†These authors contributed equally to this work*

**E-mail: songwange@nju.edu.cn, shuzhang@hku.hk*

**ABSTRACT.** Achieving fast, high-fidelity state transfer is fundamental to scalable integrated photonics and quantum information processing. While adiabatic evolution provides inherent robustness against control and fabrication imperfections, its requirement for slow driving leads to impractically long propagation distances in photonic circuits. Existing acceleration strategies, such as shortcuts to adiabaticity (STA), can dramatically shorten evolution times but generally rely on non-native auxiliary couplings or delicate Hamiltonian engineering that are difficult to implement in practice. Here we introduce evolution–pause synthesis (EPS), an interference engineered shortcut protocol that achieves fast, near-perfect state transfer strictly within the native system Hamiltonian. It achieves this by treating transient excitations as coherent resources and canceling their accumulated amplitudes via strategically interleaved pauses. By decoupling relative dynamical phase accumulation from parameter variations, EPS steers open transition trajectories into a closed loop in complex amplitude space, enabling perfect state transfer without auxiliary fields or complex parameter detours. We demonstrate this mechanism in Landau–Zener dynamics and extend it to a multilevel STIRAP process, achieving an 11.8-fold acceleration over the adiabatic baseline. Further, we experimentally validate EPS on a silicon photonic platform, realizing high-fidelity state transfer in a 16 μm footprint, a nearly tenfold reduction in device length compared with a 150 μm adiabatic reference. EPS offers a general hardware-compatible framework for fast, practical coherent control across wave and quantum platforms.

Precise and flexible manipulation of coherent states on a chip is a key capability for integrated photonics, optical information processing, and programmable wave-based devices [1-6]. In particular, complete state transfer, robust mode conversion, and high-fidelity state preparation serve as fundamental building blocks for scalable photonic circuits [7,8]. Adiabatic evolution offers a natural solution [9,10], since a system following its instantaneous eigenstate can be intrinsically insensitive to moderate fabrication or control imperfections [11]. However, this comes at the price of slow evolution. In photonic implementations, where temporal evolution is mapped onto spatial propagation along z [12], slow driving directly translates into long device lengths, limiting compactness, integration density, and scalability. To overcome this bottleneck, various acceleration strategies have been developed, among which shortcuts to adiabaticity (STA) [13,14] have attracted particular interest. Representative methods, including counterdiabatic driving [15,16] and invariant-based inverse engineering [17-19], can reproduce adiabatic-like transfer in a much shorter time or propagation length [20-23]. However, these approaches are usually Hamiltonian-centric: they accelerate evolution by introducing auxiliary interactions [24,25] or by designing delicate parameter trajectories [17,18]. Such requirements are often incompatible with realistic integrated platforms, where non-native couplings may be physically inaccessible and complex paths difficult to fabricate or control precisely [20]. As a result, practical implementation becomes more complicated, and the accelerated process may become more vulnerable to parameter noise (see Supplemental Material [28], Sec. S1 and Table S1 for a comparison of control resources and implementation requirements across representative STA schemes).

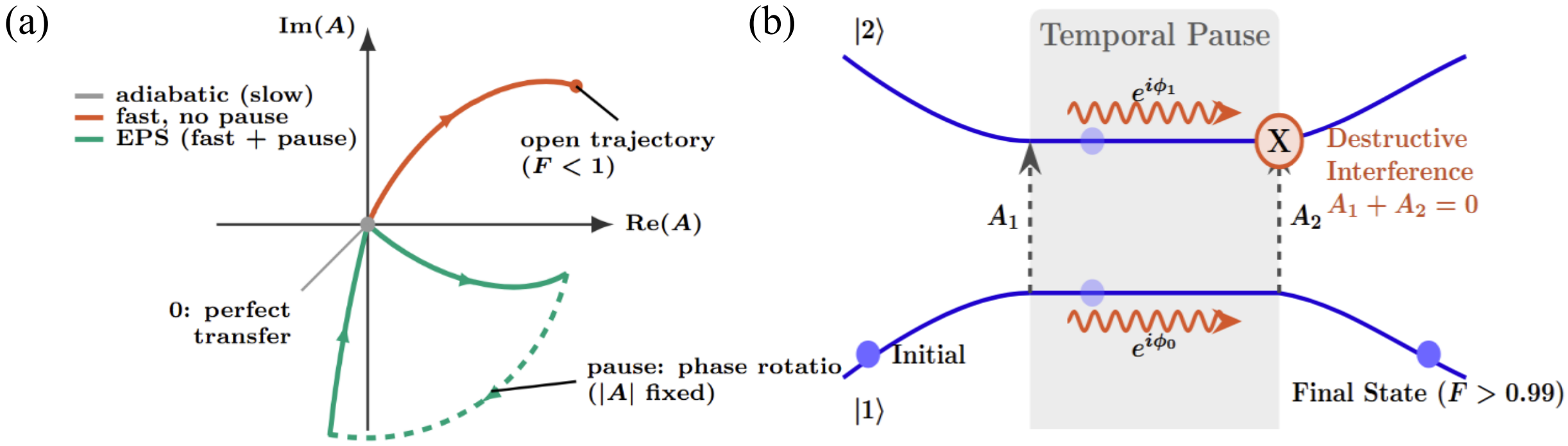

FIG. 1. Conceptual schematic of evolution-pause synthesis (EPS). (a) Complex-plane trajectory of the accumulated transition amplitude $\mathcal{A}_m(t)$ for an initially unoccupied eigenstate. Rapid driving (red) without a pause leaves $\mathcal{A}_m(T) \neq 0$, whereas the pause-induced phase rotation (green) redirects the subsequent contribution to close the trajectory at $\mathcal{A}_m(T) = 0$. (b) A pause (shaded) between two active evolution segments freezes the Hamiltonian while the eigenstate components accumulate a relative dynamical phase without interband transitions, enabling destructive interference between the transition contributions $A_1$ and $A_2$, generated before and after the pause

These limitations motivate a shift from Hamiltonian-centric control to the accumulated effects of nonadiabatic transitions. Rather than being suppressed throughout the evolution, transient excitations can be harnessed as coherent resources whose interference closes the accumulation trajectory at the final time (Fig. 1a). Achieving this closure requires independent control over the relative phases of transition contributions, which conventional protocols lack because phase accumulation remains tied to the parameter path.

In this Letter, we introduce evolution-pause synthesis (EPS), an interference-engineered shortcut that provides such phase control through strategically placed pauses along the native Hamiltonian path. Crucially, EPS preserves the geometric path of the native Hamiltonian and optimizes only its temporal parametrization by interleaving active evolution segments with pauses during which the Hamiltonian is held fixed. Thus, no auxiliary fields, non-native couplings, or complicated detours in parameter space are required. We first elucidate the underlying interference mechanism using the canonical Landau–Zener model [26], and then extend the approach to more general state-transfer processes. In a seven-level system, EPS substantially shortens the evolution time relative to the corresponding linear adiabatic process. Beyond numerical demonstrations, we experimentally implement EPS on a silicon photonic platform in a Landau–Zener configuration, achieving fidelities comparable to long-distance adiabatic protocols despite a nearly tenfold reduction in device length. Our work establishes EPS as a general and practically accessible framework for high-performance state manipulation in integrated photonics and other coherent physical systems.

To make this interference mechanism explicit, we track the complex transition amplitude accumulated in each initially unoccupied eigenstate. Writing the evolving state as $|\psi(t)\rangle = \sum_k c_k(t)|k(t)\rangle$, and adopting a fixed and continuous phase convention, we define the transition amplitude accumulated in eigenstate $m$ as

$$\mathcal{A}_m(t) \equiv \int_0^t \dot{c}_m(t')\,dt' = c_m(t) - c_m(0). \quad (1)$$

For a system initialized in eigenstate $n$, $c_m(0) = 0$ for every $m \neq n$, and hence $\mathcal{A}_m(t) = c_m(t)$. Perfect terminal transfer therefore requires the complex closure $\mathcal{A}_m(T) = 0$ for all initially unoccupied eigenstates. Instead of suppressing instantaneous transitions, this condition requires their accumulated complex amplitudes to cancel coherently at the final time. The decisive control resource therefore becomes the relative phase between transition contributions generated at different stages of evolution.

EPS provides such phase control by introducing strategically placed temporal pauses as phase shifters. During each pause, the Hamiltonian remains frozen while the system accumulates a purely dynamical phase without population transfer. By appropriately choosing the pause duration, the subsequent transition contributions interfere destructively with the previously accumulated ones, driving the total transition accumulation toward the closure condition required for perfect terminal transfer (see Fig. 1(b)). With the native Hamiltonian path fixed, EPS reduces the control problem to optimizing the pause locations and durations. Unlike Ref. [27], which selected pauses according to the sign of the instantaneous transition probability to induce adiabaticity violation during slow evolution, EPS engineers transition interference to provide a systematic framework for accelerating state transfer.

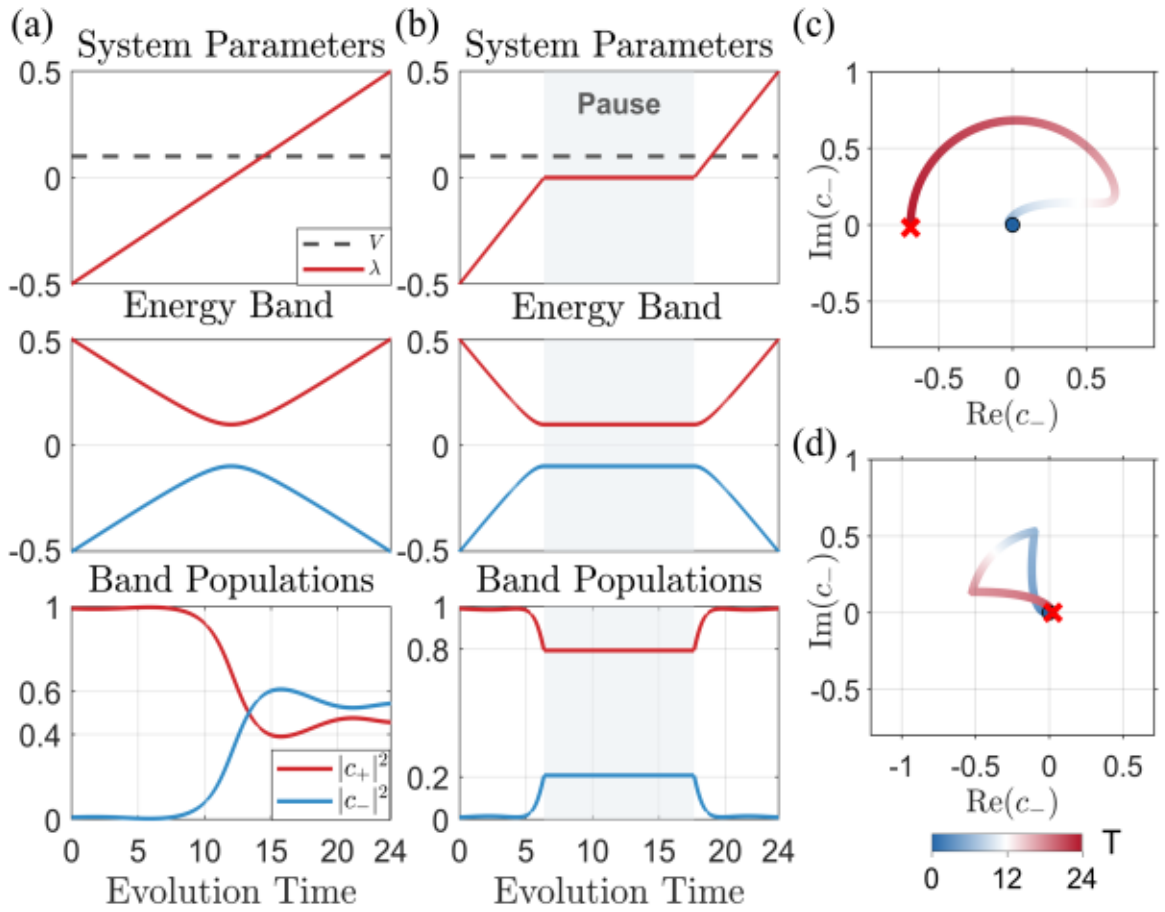

FIG. 2. Interference-engineered transition in a Landau–Zener process. (a) Rapid sweep and (b) single-pause EPS protocol for the same total evolution time. Rows show the evolution of Hamiltonian parameters $(\lambda, V)$, the energy spectrum, and the eigenstate populations $|c_{\pm}|^2$. (c, d) Complex accumulated-transition trajectories corresponding to the protocols in (a) and (b), respectively. The rapid sweep leaves an open trajectory with a residual amplitude, whereas EPS redirects the trajectory toward closure near the origin. Color encodes the evolution time.

To illustrate how independent phase control engineers transition interference, we demonstrate the EPS principle through a canonical Landau–Zener (LZ) process governed by $H(t) = \lambda(t)\sigma_z + V(t)\sigma_x$. Although the ideal LZ protocol initializes the system in an eigenstate at $\lambda \to -\infty$, we restrict the sweep to $\lambda \in$ [-0.5,0.5] for numerical and experimental feasibility, while the coupling strength is fixed at $V = 0.1$ throughout the process. Figure 2 compares a conventional rapid sweep with the EPS protocol over the same total evolution time. For clarity, we consider the simplest implementation consisting of a single temporal pause inserted at the midpoint of the sweep, exploiting the symmetry of the parameter path. As shown in Fig. 2(d), by precisely tuning the pause duration, the trajectory folds back and closes near the origin, restoring the transfer fidelity from 0.38 to above 0.99. The single pause therefore reallocates part of the fixed protocol duration from active driving to phase accumulation, enhancing the transfer fidelity without increasing the total evolution time.

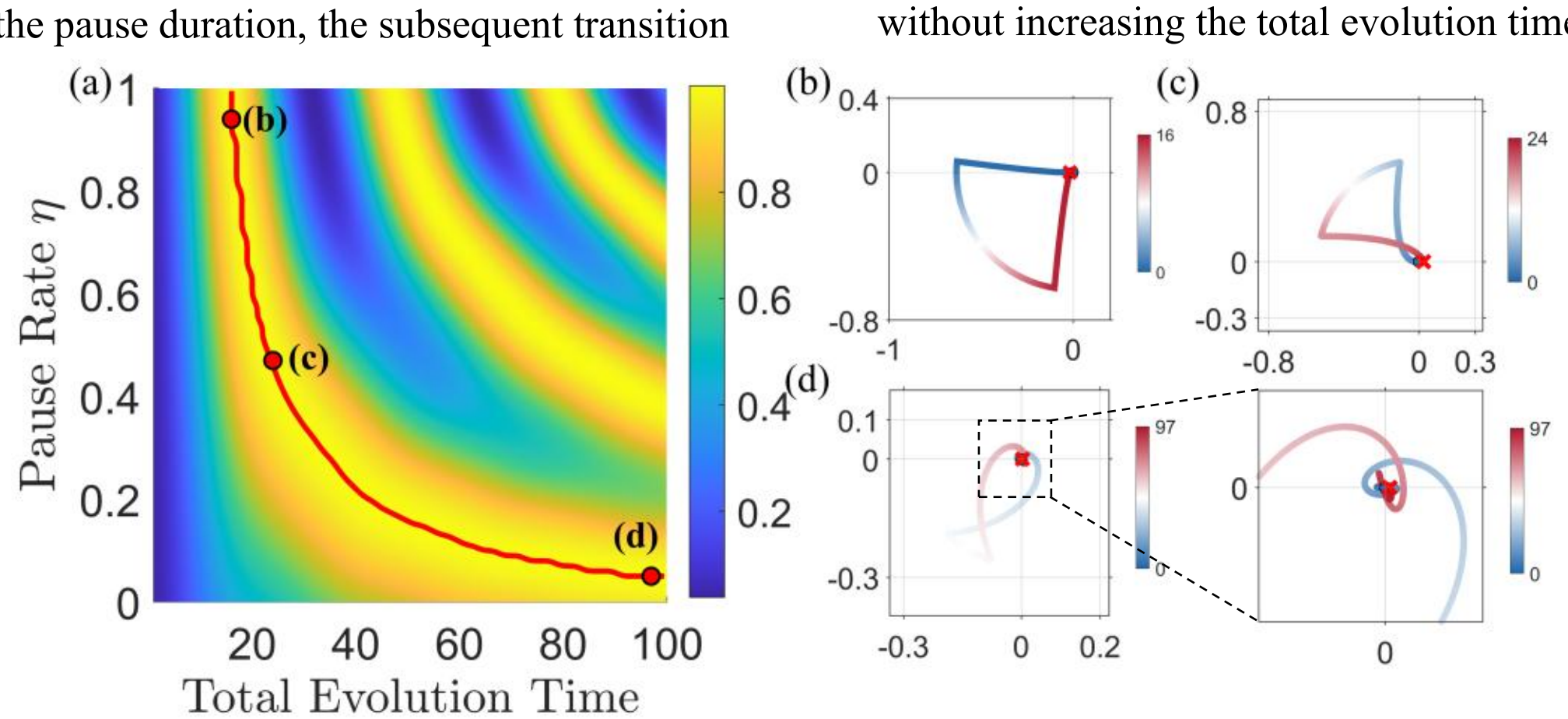

**FIG. 3. Global optimization landscape of EPS.** (a) State transfer fidelity as a function of total evolution time and pause rate $\eta$, the fraction of time occupied by the central pause. The red curve traces the optimal trajectory. (b–d) Complex transition-amplitude trajectories in the initially unoccupied instantaneous eigenstate, $c_-$, for representative protocols on the optimal trajectory at $(T, \eta) = (16, 0.94)$, $(24, 0.47)$, and $(97, 0.05)$, respectively. The lower-right trajectory shows a higher-order solution at $(T, \eta) = (97, 0.05)$, in which additional phase winding also returns the trajectory to the origin. Color encodes evolution time; circles and crosses mark the initial and final points, respectively.

To characterize the global performance of EPS, we perform a parameter sweep over the total evolution time $T$ and the pause rate $\eta$, which is defined as the fraction of time occupied by the central pause. For each combination $(T, \eta)$, we compute the final-state fidelity, yielding the landscape shown in Fig. 3. The optimal trajectory, highlighted by the red curve, traces the optimal pause rate required to achieve near-unity final-state fidelity for any given evolution time. Along this trajectory, a clear inverse scaling emerges: progressively shorter protocols require a larger fraction of independently controlled phase accumulation to preserve near-unity fidelity. The increased pause rate expands the accessible range of dynamical-phase accumulation, enhancing the capacity to redirect the stronger nonadiabatic contributions generated by faster active segments toward terminal cancellation. The representative trajectories in Figs. 3(b–d) visualize this evolution: as the active-evolution segments lengthen along the red curve, the trajectory changes from a sharply redirected excursion to increasingly winding loops around the origin. Despite this growing trajectory complexity, destructive interference consistently brings the trajectory back to its starting point, preserving near-unity fidelity. The landscape thus exposes the central control principle of EPS: evolution time is reallocated from slow adiabatic following to more efficient phase engineering, enabling acceleration without loss of fidelity. Since EPS introduces no auxiliary Hamiltonian terms, it avoids additional control-error channels associated with implementing non-native couplings. Numerical tests further confirm that high fidelity is retained under coupling and detuning perturbations (see Supplemental Material [28], Sec. S2).

To establish the experimental feasibility of EPS, we implement the protocol on the silicon-on-insulator (SOI) photonic waveguide platform. Under the paraxial approximation, light propagation in two evanescently coupled single-mode waveguides is described by coupled-mode theory (CMT) [29]. The amplitude vector $\mathbf{a}(z) = [a_1(z), a_2(z)]^{\mathrm{T}}$ satisfies

$$i\frac{d\mathbf{a}}{dz} = H_{\mathrm{CMT}}(z)\mathbf{a},\ H_{\mathrm{CMT}}(z) = \begin{pmatrix} \beta_1(z) & \kappa(z) \\ \kappa(z) & \beta_2(z) \end{pmatrix}. \quad (2)$$

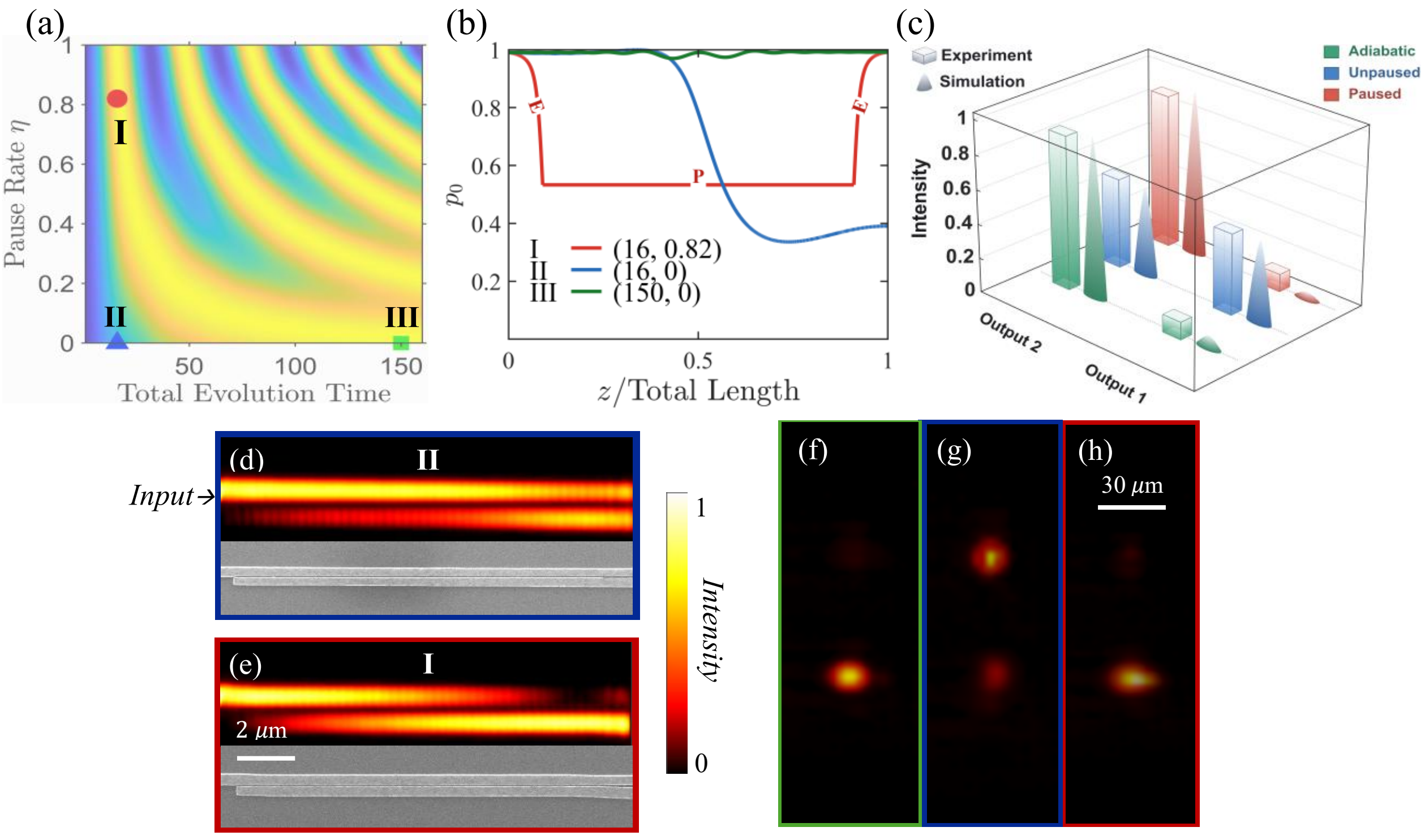

FIG. 4. Experimental realization of EPS on a silicon photonic platform. (a) Final-state-fidelity landscape with markers I–III denoting the 16 μm EPS device ($\eta = 0.82$), 16 μm no-pause device, and 150 μm adiabatic reference. (b) Simulated population $p_0$ of desired instantaneous eigenstate along the normalized propagation distance $z/L$ for I–III. E–P–E denotes the evolution–pause–evolution sequence, with the Hamiltonian varying during E and fixed during P. (c) Experimental (cuboids) and simulated (cones) unit-normalized output intensities for the adiabatic (green), no-pause (blue), and EPS (red) devices. (d, e) Simulated propagation-intensity distributions (top) and corresponding SEM images (bottom) of the no-pause and EPS devices. (f–h) Corresponding near-infrared CCD output-intensity distributions measured at 1550 nm.

Here, $\beta_{1,2}(z)$ are the propagation constants of the individual waveguides and $\kappa(z)$ is their evanescent coupling coefficient. After removing the common propagation constant, the Hamiltonian can be written as

$$H_{\mathrm{CMT}}(z) = \begin{pmatrix} \Delta\beta(z)/2 & \kappa(z) \\ \kappa(z) & -\Delta\beta(z)/2 \end{pmatrix}, \quad (3)$$

where $\Delta\beta = \beta_1 - \beta_2$. This expression takes the same form as the Landau–Zener Hamiltonian introduced above, with the propagation-constant difference replacing the energy detuning and the evanescent coupling providing the off-diagonal term. In the waveguide implementation, the propagation constants are controlled by the respective waveguide widths, whereas the coupling coefficient is controlled by the inter-waveguide gap. Under this isomorphism, the strategic temporal pauses are realized as longitudinally invariant waveguide segments, where both the waveguide widths and the inter-waveguide gap remain constant along the propagation direction $z$.

Experimentally, we compare three coupler configurations, labelled I–III in Fig. 4. Device I is a 16 $\mu$m-long EPS coupler with a pause rate of $\eta = 0.82$, corresponding to a 13.12 $\mu$m pause at the midpoint flanked by two 1.44 $\mu$m evolution segments. Device II is a 16 $\mu$m-long diabatic coupler without a pause ($\eta = 0$), providing a direct comparison with the EPS device at the same footprint. Device III is a 150 $\mu$m-long adiabatic coupler without a pause, serving as the reference device. The device geometries were inversely designed and validated through full-wave simulations using COMSOL Multiphysics, ensuring accurate implementation of the target Hamiltonian [30] (see Supplemental Material [28]). Details of the device fabrication and experimental characterization are provided in the Supplemental Material [28], Sec. S5. As expected, directly compressing the conventional adiabatic device results in severe fidelity degradation due to uncontrolled nonadiabatic excitations. By contrast, the EPS device achieves fidelity comparable to that of the long adiabatic reference within a nearly tenfold shorter footprint, experimentally demonstrating pause-induced phase control along the native Hamiltonian path. The measured output-state fidelities and device-to-device statistics are reported in the Supplemental Material [28], Table S2. Full-wave simulations further indicate that the slope discontinuities at the evolution–pause boundaries do not introduce appreciable additional scattering under the simulated conditions (see Supplemental Material [28], Fig. S4).

To test whether EPS remains effective in a more complex multilevel setting, we consider a seven-level generalization of stimulated Raman adiabatic passage (STIRAP). The system forms a nearest-neighbor coupled chain and supports a zero-eigenvalue dark state connecting the two endpoint states [31,32]. In the diabatic-state basis, the amplitude vector $\mathbf{c}(t) = [c_1(t), c_2(t), \ldots, c_7(t)]^{\mathrm{T}}$, which obeys

$$i\frac{d\mathbf{c}}{dt} = H_7(t)\mathbf{c}, \quad (4)$$

where the coupling Hamiltonian is

$$H_7(t) = \begin{pmatrix} 0 & \Omega_{\mathrm{o}} & 0 & 0 & 0 & 0 & 0 \\ \Omega_{\mathrm{o}} & 0 & \Omega_{\mathrm{e}} & 0 & 0 & 0 & 0 \\ 0 & \Omega_{\mathrm{e}} & 0 & \Omega_{\mathrm{o}} & 0 & 0 & 0 \\ 0 & 0 & \Omega_{\mathrm{o}} & 0 & \Omega_{\mathrm{e}} & 0 & 0 \\ 0 & 0 & 0 & \Omega_{\mathrm{e}} & 0 & \Omega_{\mathrm{o}} & 0 \\ 0 & 0 & 0 & 0 & \Omega_{\mathrm{o}} & 0 & \Omega_{\mathrm{e}} \\ 0 & 0 & 0 & 0 & 0 & \Omega_{\mathrm{e}} & 0 \end{pmatrix}, \quad (5)$$

where $\Omega_{\mathrm{o}}(t)$ and $\Omega_{\mathrm{e}}(t)$ denote the odd- and even-indexed nearest-neighbor couplings respectively, in direct analogy to the pump and Stokes pulses in conventional STIRAP.

To more fully assess EPS in this multilevel system, we extend the single-pause scheme to a multi-pause protocol. Under identical boundary conditions and coupling ranges, we compare EPS with a conventional linear protocol that satisfies the strict adiabatic criterion. The Hamiltonian path is parametrized by a scalar path coordinate $s \in [0,1]$, such that $\mathbf{\Omega}(s) = (\Omega_o, \Omega_e) = (s, 1-s)$, with boundary conditions $\mathbf{\Omega}(0) = (0,1)$ and $\mathbf{\Omega}(T) = (1,0)$. The path evolution is partitioned into active and pause segments, with its local velocity restricted to the binary choice: $\dot{s}(t) \in \{\mathrm{v}_0, 0\}$. A gradient-descent algorithm is then used to optimize the distribution and duration of these pauses to maximize end-state fidelity at each total evolution time T.

Here, the linear ramp provides a stringent test for isolating the acceleration enabled by EPS. Under the normalized coupling amplitude, the adiabatic linear protocol requires an evolution time of $\mathrm{T} = 118.63$ to maintain an instantaneous dark-state fidelity above 0.99 throughout the evolution. As a shortcut protocol, EPS is not required to follow the instantaneous dark state and is instead evaluated by its terminal transfer fidelity. EPS reaches a terminal fidelity above 0.99 at $\mathrm{T} = 10.09$, corresponding to a 91.49% reduction in evolution time, or an 11.76-fold acceleration. The seven-level dark-state construction, numerical optimization procedure, optimized coupling trajectories, and corresponding instantaneous fidelities are provided in the Supplemental

Material [28], Sec. S3. Importantly, this improvement is achieved without modifying the geometric path of the Hamiltonian or introducing auxiliary couplings. These results demonstrate that the interference mechanism underlying EPS remains highly effective beyond two-level dynamics.

In conclusion, we have introduced an interference-engineered shortcut to perfect state transfer that operates entirely within the native Hamiltonian manifold. Rather than suppressing nonadiabatic transitions throughout evolution, the protocol uses evolution-pause synthesis (EPS) to control the relative phases of transition amplitudes, causing them to cancel destructively. In the Landau–Zener model, a single pause inserted at the midpoint restored near-perfect transfer. We then implemented this principle in silicon waveguides, achieving performance comparable to that of the adiabatic reference in a device nearly tenfold shorter. Extending EPS to a seven-level STIRAP system further demonstrates that the acceleration enabled by pause-induced interference persists in multilevel dynamics.

More broadly, the interference-engineered shortcut shifts the design of coherent evolution from engineering elaborate Hamiltonian trajectories to shaping the accumulated transition amplitudes, and identifies relative phases as important control resources. Because it requires only pauses along the native evolution path, the protocol can be readily extended to complex systems governed by time-dependent Hamiltonians, provided that the Hamiltonian can be temporarily held fixed to control relative dynamical phases. EPS therefore offers a general route to compact and high-fidelity shortcuts across photonic, atomic, molecular, and other quantum platforms.

### Acknowledgements

The authors acknowledge the financial support from The National Key R&D Program of China (No. 2023YFA1407700), National Natural Science Foundation of China (No. 12522421), Quantum Science Center of Guangdong-Hong Kong-Macau Great Bay Area, New Cornerstone Science Foundation, the Hong Kong Research Grant Council (STG3/E-704/23-N, AoE/P-701/20, 17315522), and the Guangdong Provincial Quantum Science Strategic Initiative (No. GDZX2204004 and No. GDZX2304001).